\documentclass[prl,twocolumn,showpacs,preprintnumbers,amsmath,amssymb]{revtex4}

\usepackage{graphicx}
\usepackage{rotating}
\usepackage{dcolumn}
\usepackage{bm}

\begin{document}

\preprint{cond-mat/somewhere}

\title{Control of Interaction-Induced Dephasing of Bloch Oscillations} 

\author{M. Gustavsson, E. Haller, M.~J. Mark, J.~G. Danzl, G. Rojas-Kopeinig, H.-C. N\"{a}gerl}
\affiliation{ $^1$Institut f\"{u}r Experimentalphysik und
Forschungszentrum f\"{u}r Quantenphysik, Universit\"{a}t
Innsbruck,  6020 Innsbruck, Austria}
%

\date{\today}

\begin{abstract}
We report on the control of interaction-induced dephasing of Bloch
oscillations for an atomic Bose-Einstein condensate in an optical
lattice under the influence of gravity. When tuning the strength
of the interaction towards zero by means of a Feshbach resonance,
the dephasing time is increased from a few to more than twenty
thousand Bloch oscillation periods. We quantify the dephasing in
terms of the width of the quasi-momentum distribution and measure
its dependence on time for different values of the scattering
length. Minimizing the dephasing allows us to realize a BEC-based
atom interferometer in the non-interacting limit. We use it for a
precise determination of a zero-crossing for the atomic scattering
length and to observe collapse and revivals of Bloch oscillations
when the atomic sample is subject to a spatial force gradient.
\end{abstract}



\maketitle

Ultracold atomic systems have initiated a revolution in the field
of precision measurements. Laser cooled thermal samples are used
for ultra-high resolution laser spectroscopy \cite{Diddams2004},
they are at the heart of modern atomic fountain clocks
\cite{Bize2005,Boyd2007}, and they allow the realization of
matter-wave interferometers for high-precision inertial sensing
\cite{Peters1999} and high-precision determination of fundamental
constants \cite{Clade2006}. Atomic Bose-Einstein condensates
(BEC), the matter-wave analoga to the laser, combine high
brightness with narrow spatial and momentum spread. In general,
the resolution is limited only by the quantum mechanical
uncertainty principle, and BECs could thus serve as ideal sources
for precision measurements and in particular for matter wave
interferometers \cite{Gupta2002}. Atom-atom interactions, however,
have to be taken into account, as they lead to collisional
dephasing and give rise to density dependent mean-field shifts in
the interferometric signal. It is thus advisable to either operate
a BEC-based atom interferometer in the dilute density limit,
possibly sacrificing a high signal-to-noise ratio, or to find ways
of reducing or even nulling the strength of the interaction
altogether. Precisely the latter is feasible in the vicinity of
magnetically induced Feshbach resonances where the atomic s-wave
scattering length and hence the strength of the atom-atom contact
interaction go through a zero crossing \cite{Koehler2006}. It is
thus possible to experimentally investigate the reduction and even
disappearance of interaction-induced effects on the
interferometric signal as the scattering length is tuned towards
zero by means of an externally controlled magnetic field.

A paradigm atom interferometric effect is the well-known
phenomenon of Bloch oscillations \cite{Dahan1996}. Bloch
oscillations for the mean quasi-momentum are the result of single
atom interference as the atomic wavepacket, subject to a constant
force, is Bragg reflected in the presence of a periodic optical
lattice potential. They have been observed for ultracold thermal
samples \cite{Dahan1996,Battesti2004,Clade2006,Ferrari2006}, for
atoms in interacting BECs \cite{Morsch2001,Roati2004}, and for
ensembles of non-interacting quantum-degenerate fermions
\cite{Roati2004}. For the case of the interacting BEC, strong
dephasing is found as evidenced by a rapid broadening and apparent
smearing out of the momentum distribution in the first Brillouin
zone, limiting the observation of Bloch oscillations to a few
cycles for typical atomic densities in a BEC. In addition, the
measured initial width of the momentum distribution is comparable
to the extent of the Brillouin zone, as interaction energy is
converted into kinetic energy upon release of the BEC from the
lattice potential, thus greatly reducing the contrast of the
oscillations \cite{Roati2004}.

In this Letter, we report on the control of interaction induced
dephasing of Bloch oscillations for a BEC in a vertically oriented
optical lattice under the influence of gravity. Control is
obtained by means of a zero crossing for the atomic $s$-wave
scattering length $a$. We observe the transition from an
interacting BEC to a non-interacting BEC by measuring the rate of
dephasing, given by the change of the width of the momentum
distribution, as a function of $a$. We identify a clear minimum
for the dephasing which we associate with the zero crossing for
$a$. At the minimum more than $2\!\times\!10^4$ oscillations can
be observed with high contrast, and the zero crossing can be
determined with high precision. For our measurements at non-zero
scattering length, we greatly reduce broadening of the momentum
distribution by rapidly switching the interaction strength to zero
upon release from the lattice potential. Our measurements indicate
that BECs can indeed be used as a source for precision atom
interferometry, as effects of the interaction can be greatly
reduced. For a non-interacting BEC, we intentionally induce
dephasing by means of a weak optical force gradient and observe
collapse and revivals of Bloch oscillations.
\begin{figure}
\includegraphics[width=0.48\textwidth]{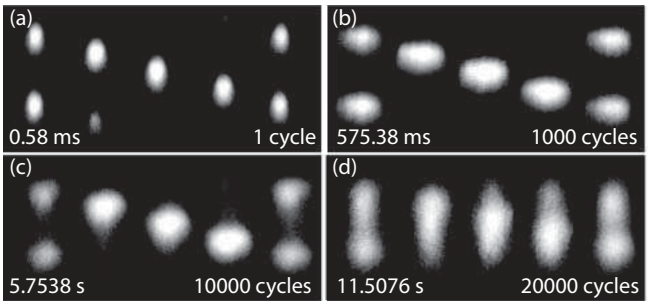}
\caption{Long-lived Bloch oscillations for a non-interacting BEC
with Cs atoms in the vertical lattice under the influence of
gravity. Each picture shows one Bloch cycle in successive
time-of-flight absorption images corresponding to the momentum
distribution at the time of release from the lattice. Displayed
are the first (a), the $ 1000^{\text{th}}$ (b), the $
10000^{\text{th}}$ (c), and the $ 20000^{\text{th}}$ (d) Bloch
cycle for minimal interaction near the zero crossing for the
scattering length.} \label{Fig1}
\end{figure}

The starting point for our experiments is an essentially pure BEC
with typically $1 \! \times \! 10^5$ Cs atoms in the $|F\!=\!3,
m_F\!=\!3\rangle$ hyperfine ground state sublevel confined in a
crossed-beam dipole trap generated by one vertically (L$_1$, with
$1/e^2$-beam diameter 256 $\mu$m) and one more tightly focused
horizontally (L$_2$, with diameter 84 $\mu$m) propagating laser
beam at a wavelength near 1064\,nm. We support the optical
trapping by magnetic levitation against gravity \cite{Weber2003}.
For BEC preparation, we basically follow the procedure described
in Ref.\,\cite{Weber2003,Kraemer2004}. The strength of the
interaction can be tuned by means of a broad Fesh\-bach resonance,
which causes a zero crossing for the scattering length $a$ near an
offset magnetic field value of $17$ G with a slope of
$61$\,a$_0/$G \cite{Julienne}. Here, a$_0$ denotes Bohr's radius.
The lattice potential is generated by a vertically oriented
standing laser wave generated by retro-reflection, co-linear with
L$_1$, but with much larger diameter of 580 $\mu$m. This allows
independent control of lattice depth and radial (i.e. horizontal)
confinement. The light comes from a home-built single-mode fiber
amplifier \cite{Liem2003} seeded with highly-stable light at
$\lambda = 1064.4946(1)$ nm. We turn on the optical lattice
potential exponentially to a depth of $7.9 \ E_R$ within $ 1000 $
ms, where $E_R=h^2/(2 m \lambda^2) = k_B \! \times \! 64 \, $nK is
the photon recoil energy and $m$ is the mass of the Cs atom. The
slow ramp assures that the BEC is adiabatically loaded into the
lowest Bloch band of the lattice. We load between 40 to 65 lattice
sites, depending on the initial vertical extent of the BEC. We
then reduce the power in L$_2$ to zero within $ 300 \, \mu$s.
Subsequently, the magnetic field gradient needed for levitation is
ramped down and a bias magnetic field is tuned to the desired
value within $ 100 \, \mu$s. For the present experiments, we
adjust $a$ in the range from $-2$ to $300 \, a_0$ with magnetic
bias fields from 17 to 23 G. We control the average bias field to
about 1 mG. The confinement of the BEC in the lattice as given by
L$_1$ gives horizontal trapping frequencies in the range of $5$ to
$10$ Hz. We then let the atoms evolve in the lattice under the
influence of the gravitational force for variable hold time $ T $.
Finally, we switch off the horizontal confinement and ramp the
lattice depth adiabatically to zero within $300 \, \mu$s to
measure the momentum distribution by the standard time-of-flight
technique, taking an absorption picture on a CCD camera. For some
of the data we turn on the magnetic levitation field to allow for
longer expansion times up to 100 ms. To minimize broadening of the
distribution as a result of interaction we switch the scattering
length to zero during the release and the initial time-of-flight.

We observe persistent Bloch oscillations when minimizing the
effect of interactions at a magnetic field value of $17.12$ G (see
below). Fig.\,\ref{Fig1} (a)-(d) show the evolution of the
momentum distribution during the first, the $ 1000^{\text{th}}$,
the $ 10000^{\text{th}}$, and the $ 20000^{\text{th}}$ Bloch
cycle. Initially, the momentum distribution exhibits narrow peaks.
Their full width $\Delta p$ \cite{width} is as narrow as about
$0.15 \ \hbar k$, where $k\!=\!2\pi/\lambda$. Very little
broadening along the vertical direction is seen after the first
1000 cycles. Initial excitation of horizontal motion as a result
of ramping the power in L$_2$ and switching the scattering length
leads to some horizontal spreading. After 20000 cycles the
distribution has started to spread out noticeably along the
vertical direction.
\begin{figure}
\includegraphics[width=0.48\textwidth]{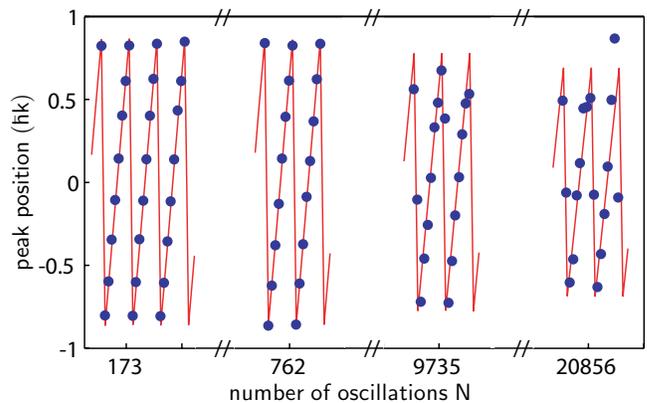}
\caption{(color online). Position of the strongest peak in the
momentum distribution as a function of the number $N$ of Bloch
oscillations (dots). More than 20000 cycles can be followed with
high contrast. A fit to the data (solid curve) yields a Bloch
period of $0.5753807(5)$ ms.} \label{Fig2}
\end{figure}

Fig.\,\ref{Fig2} highlights the high number of Bloch oscillations,
which we can observe for the case of minimal interaction strength.
It shows how the strongest peak of the momentum distribution
cycles through the first Brillouin zone with the typical sawtooth
behavior \cite{Dahan1996}. More than 20000 cycles can easily be
followed. From a fit to the data we determine the Bloch period to
$0.5753807(5)$ ms. Assuming that no additional forces act on the
sample, the local gravitational constant is $ g=9.804450(9) \,
$m/s$^2$.
\begin{figure}
\includegraphics[width=0.48\textwidth]{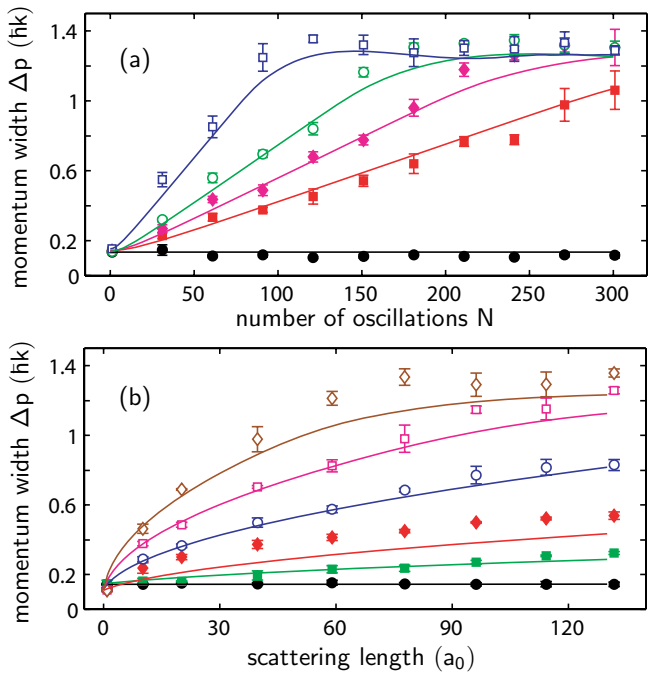}
\caption{(color online). Width $\Delta p$ of the momentum
distribution for different interaction strengths. (a) Evolution of
$\Delta p$ as a function of the number $N$ of Bloch cycles for
different values of the scattering length ($a\!=\!0, 25, 50, 100,
$ and $ 300 \, a_0$ from bottom (full circles) to top (full
diamonds). The solid curves are derived from a numerical model
calculation, see text. (b) Width $\Delta p$ for a fixed number of
cycles $N\!=\!1 $ (full circles), $25$ (full squares), $50$ (full
diamonds), $100$ (open circles), $150$ (open squares), and $200$
(open diamonds) as a function of scattering length. The solid line
represents the model calculation. All error bars correspond to
$\pm$ one standard deviation resulting from 7 measurements. The
data and the simulations correspond to the following parameters:
lattice depth: $7.9 \ E_R$, scattering length during lattice
loading: $210 \ a_0$, trapping frequencies in L$_1$ and L$_2$: 10
and 8 Hz, atom number in the BEC: $5\!\times\!10^4$.} \label{Fig3}
\end{figure}

In order to quantify the dephasing of Bloch oscillations we
determine for each Bloch period the width $\Delta p$ of the
momentum distribution at the instant in time when the peak of the
distribution is centered at zero momentum, i.e.~for the central
picture of each series shown in Fig.\,\ref{Fig1}. Fig.\,\ref{Fig3}
(a) displays $\Delta p$ up to the $300^\text{th}$ Bloch cycle for
different interaction strengths ranging from $0$ to $300 \, a_0$.
For minimal interaction strength ($a \! \approx \! 0 \, a_0$), we
see no broadening of the distribution. Broadening can clearly be
seen for $a \!=\! 25 \, a_0$, and the rate of broadening then
increases with increasing interaction strength. For $a \! \ge \!
50 \, a_0$ the width $\Delta p$ saturates within the chosen
observation time to a value of about $1.3 \ \hbar k$ as the
momentum distribution completely fills the first Brillouin zone
\cite{structure}. To a good approximation, we find that $\Delta p$
initially increases linearly with time. In Fig.\,\ref{Fig3} (b) we
plot $\Delta p$ as a function of interaction strength for various
fixed numbers of Bloch cycles. $\Delta p$ appears to scale with
the square root of the interaction strength. Both observations
agree well with a simple model for the dephasing of Bloch
oscillations, which predicts $ \Delta p \! \propto \! \sqrt{a} \!
\times \! T$ \cite{Witthaut2005} for sufficiently short times $T$.
In order to verify this model, we have performed numerical
calculations solving the one-dimensional Gross-Pitaevskii equation
in the presence of an optical lattice under the influence of
gravity for the typical parameters of our experiment according to
the method detailed in Ref.~\cite{Smerzi2003}. Via Fourier
transform of the spatial wave function we determine the momentum
distribution and its width. As shown in Fig.\,\ref{Fig3} (solid
lines) we find very good agreement with our measurements with no
adjustable parameters when we add a constant offset of $0.1 \
\hbar k$ to all the numerical curves. This offset takes into
account residual interactions during release from the lattice as a
result of the finite magnetic switching speed, which leads to some
artificial broadening of the distribution. We attribute the
systematic discrepancy for the $N=50$ data in Fig.\,\ref{Fig3} (b)
to the horizontal motion which leads to modulations in the density
that adds a modulation onto $ \Delta p$ also seen in
Fig.\,\ref{Fig3} (a).
\begin{figure}
\includegraphics[width=0.48\textwidth]{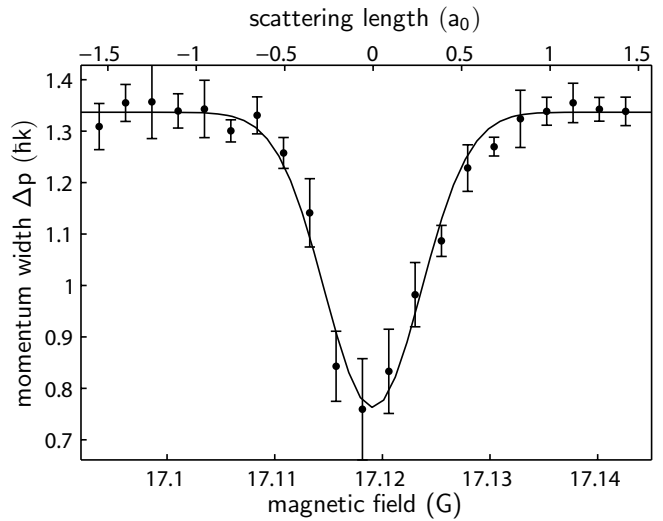}
\caption{Broadening of the momentum distribution as a result of
$6951$ Bloch oscillations near the zero crossing for the
scattering length. The width $\Delta p$ is plotted as a function
of magnetic field (dots). The solid line is a Gaussian fit with a
rms-width of $4.5$ mG. The fit is centered at $17.119(2)$ G. The
zero for the scattering length scale on top was chosen to agree
with this value.} \label{Fig4}
\end{figure}

To find the value for the magnetic field that gives minimal
broadening we measure $\Delta p$ after $6951$ cycles in the
vicinity of the crossing. Fig.\,\ref{Fig4} plots $\Delta p$ as a
function of magnetic field. It shows a clear minimum, which we
expect to correspond to the zero crossing for the scattering
length. From a Gaussian fit we determine the center position of
the minimum to be at $17.119(2)$ G. The one-sigma error takes into
account our statistical error in magnetic field calibration. To
our knowledge, this is by far the most precise determination of a
Ramsauer-Townsend minimum \cite{RamsauerTownsend} to date in
ultracold atom scattering. We believe that our measurements are
limited by the ambient magnetic field noise, leading to a finite
width for the distribution of the scattering length. In fact, a
reduction of the atomic density gives longer decay times for the
Bloch oscillations. Note that in the scattering length regime
considered here the effect of the (magnetic) dipole-dipole
interaction \cite{Giovanazzi2002} should start to play a role.
\begin{figure}
\includegraphics[width=0.48\textwidth]{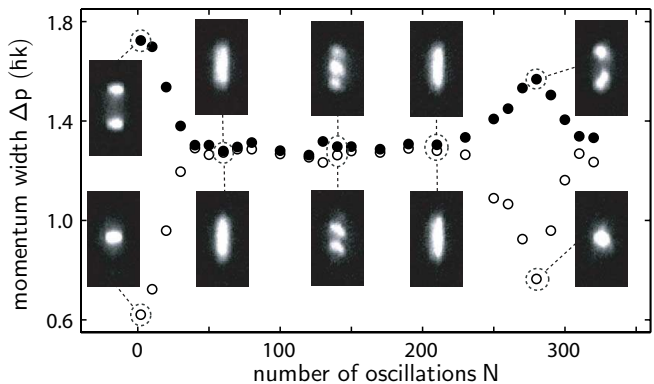}
\caption{Collapse and revival of Bloch oscillations for the case
of a non-interacting BEC with a vertical force gradient. For two
cycle phases separated by $\pi$, the width $\Delta p$ is plotted
as a function of the number $N$ of Bloch cycles. For selected
cycles ($N=1, 70, 140, 210,$ and $280$) two absorption images
corresponding to the two cycle phases are shown. } \label{Fig5}
\end{figure}

Our capability to observe Bloch oscillations on extended time
scales without interaction-induced dephasing allows us to study
the effect of deliberately imposed dephasing. For this we apply a
linear force gradient $ \nabla\!F $ corresponding to harmonic
trapping at $\nu\!=\!40(1)$ Hz along the vertical direction by
turning on L$_2$ during the hold time. Fig.\,\ref{Fig5} shows the
widths $\Delta p$ for two cycle phases separated by $\pi$
initially corresponding to the single resp. double-peaked
distribution as a function of the number $N$ of Bloch cycles. Both
widths rapidly increase resp.~decrease to the same value of $1.3 \
\hbar k$ within about $N\!=\!30$ oscillations. Here the ensemble
is dephased. It then remains dephased for about 200 cycles.
Partial rephasing at intermediate times not reflected in the
widths can be seen from the absorption images. Revival of the
oscillations \cite{Ponomarev2006} happens around $N\!=\!280$ when
the values for both widths separate again
\cite{collapse_and_revival}. This number agrees well with the
expected value of $N_\text{rev}\!=\!292(15)$ given by
$N_\text{rev}\!=\!F_\text{grav}/(\nabla\!F d)\!=\!m g/(m \omega^2
d)$, where $F_\text{grav}$ is the gravitational force,
$\omega\!=\!2\pi\nu$, and $d\!=\!\lambda/2$ is the lattice
spacing. Subsequently, the widths collapse again to the common
value. In further measurements we see up to four collapses and
revivals.

In summary, we have demonstrated the control of
interaction-induced dephasing near a zero-crossing for the
scattering length. On the crossing, we have realized a
non-interacting BEC, which allows us to observe more than $20000$
Bloch cycles, indicating a matter wave coherence time of more than
$10$ s. The broadening of the momentum distribution agrees well
with results from theoretical models. We believe that the number
of observable Bloch cycles is limited by residual interactions as
a result of magnetic field noise. Our results open up exciting new
avenues for precision measurements with quantum degenerate gases.
For example, it is now possible to perform sensitive measurements
of forces on short length scales, such as the Casimir-Polder force
near a dielectric surface \cite{Carusotto2005}. Future
experimental work can now address the nature of the dephasing
\cite{Buchleitner2003} by studying structure in the momentum
distribution.

A similar experiment on long-lasting Bloch oscillations and
control of the interaction strength has recently been performed
with a BEC of $^{39}$K atoms at LENS, Italy. We thank A. Daley for
theoretical support and for help with setting up the numerical
calculations and A. Buchleitner and his group for useful
discussions. We are grateful to A. Liem and H. Zellmer for
valuable assistance in setting up the 1064-nm fiber amplifier
system. We acknowledge contributions by P. Unterwaditzer and T.
Flir during the early stages of the experiment. We are indebted to
R. Grimm for generous support and gratefully acknowledge funding
by the Austrian Ministry of Science and Research (BMWF) and the
Austrian Science Fund (FWF) in form of a START prize grant.


\begin{references}
\bibitem{width} We define the full width $ \Delta p $ to be the
root-mean-square (rms) diameter of the distribution.
\bibitem{structure} The momentum distribution
of a fully dephased interacting ensemble carries high-contrast
substructure which will be discussed in a forthcoming publication.
\bibitem{collapse_and_revival} Note that the revived Bloch cycles are
subject to a phase shift, which depends on the vertical location
of the harmonic trap minimum with respect to the lattice minima.
\end{references}
\end{document}